\documentclass[aps,PRResearch,reprint,groupedaddress]{revtex4-1}

\usepackage{soul,color}

\usepackage{amsmath,bm}
\usepackage{graphicx}
\usepackage[raggedright,hang,nooneline]{subfigure}
\usepackage{float}

\newcommand{\pvec}[1]{\vec{#1}\mkern2mu\vphantom{#1}}

\begin{document}


\title{Stability and dynamics of optically levitated dielectric disks in a Gaussian standing wave beyond the harmonic approximation}


\author{T. Seberson$^{1}$ and F. Robicheaux$^{1,2}$}
\affiliation{$^{1}$Department of Physics and Astronomy, Purdue University, West Lafayette, Indiana 47907, USA}
\affiliation{$^{2}$Purdue Quantum Science and Engineering Institute, Purdue University, West Lafayette, Indiana 47907, USA}


\date{\today}

\begin{abstract}
Forces and torques exerted on dielectric disks trapped in a Gaussian standing wave are analyzed theoretically for disks of radius $2~\mu\text{m}$ with index of refraction $n=1.45$ and $n=2.0$ as well as disks of radius 200 nm with $n=1.45$. Calculations of the forces and torques were conducted both analytically and numerically using a discrete-dipole approximation method. Besides harmonic terms, third order ro-translational coupling terms in the potential energy can be significant and a necessary consideration when describing the dynamics of disks outside of the Rayleigh limit. The coupling terms are a result of the finite extension of the disk coupling to both the Gaussian and standing wave geometry of the beam. The resulting dynamics of the degrees of freedom most affected by the coupling terms exhibit several sidebands as evidenced in the power spectral densities. Simulations show that for Gaussian beam waists of $2-4~\mu\text{m}$ the disk remains stably trapped. 
\end{abstract}


\maketitle


\section{Introduction}
The choice of particle used in levitated optomechanics is an important factor that depends on the goal of application. The most widely used particle in the field is a silica sphere with radius small compared to the wavelength. The dynamics of spheres trapped in cavities and focused laser beams are well understood and used for cooling to the motional ground state as well as force sensing \cite{Delic2020,Ranjit_2016,Millen_2020,PhysRevA.101.053835}. This is owing to the simple harmonic translational and free rotational dynamics making it an ideal system to handle for both experimentalists and theorists. Particles with decreased particle symmetry allow rotational degrees of freedom to enter into the potential energy. A nanorod has large differences in moments of inertia and polarizability which allows rotations to be described as decoupled librations about the laser polarization axis. The motion of nanodumbells or generally anisotropic materials requires rigid-body dynamics since these particles have moments of inertia of similar magnitude \cite{Seberson2019,Rashid2018}. Increasing the size of the particle relative to the wavelength of the laser further complicates the motion for any particle shape \cite{Neto_2000}. Still, terms necessary to describe nanorods and nanodumbbells have been investigated and the motion is also well understood \cite{Kuhn:17,Stickler_2018,Kuhn2017a,bang20205d}. 

Dielectric disks also have a relatively simple shape, but have not seen as much attention as other particle geometries. Several studies point to thin nanodisk scattering being more realistically described in a Rayleigh-Gans rather than a Rayleigh approximation for index of refraction $n\sim1$ \cite{Willis:87,LeVine:83,doi:10.1063/1.326257}. This generally leads to an orientational dependent shape function in the form of a Bessel function. From studies investigating the applications of disks for various purposes, it is unclear whether there is consensus on the necessity of including the shape function or other non-harmonic terms in the dynamics \cite{Stickler2016a,Arvanitaki2013,Chang_2012}. There are few experimental studies involving disks, however two such studies suggest terms of higher order may be necessary for describing the motion \cite{PhysRevE.68.051404,PhysRevLett.89.108303}. 

In this paper it is shown that higher order terms of at least third order in the potential energy are necessary for describing the dynamics of disks outside the Rayleigh regime in a Gaussian standing wave. While a focused Gaussian traveling wave is the most common trap, the large radiation pressure exerted on disks raises the concern for instability. In a Gaussian standing wave (e.g. driven cavity) axial confinement is much stronger and the axial radiation pressure is absent. A disk experiences restoring forces in all three translational degrees of freedom and torques in two rotational degrees of freedom. Similar to rods and nanodumbells, the rotation about the disk's symmetry axis is unaffected by light coupling and is a constant of the motion. Focus is given to the effects due to the third order terms which provide unique ro-translational couplings that have not yet been discussed in levitated optomechanics. The coupling terms are a result of the finite extension of the disk coupling to both the Gaussian and standing wave geometry of the beam. Inclusion of the coupling terms results in dynamics with several different modes of oscillation for each degree of freedom which are evident in the power spectral density. Simulations show no evidence of instability. 

An analytical as well as numerical approach using a discrete-dipole approximation method is used to identify the forces and torques on disks of radius $2~\mu\text{m}$ with index of refraction $n=1.45$ and $n=2.0$ as well as disks of radius 200 nm with $n=1.45$. The Gaussian standing wave is constructed with a wavelength $\lambda = 850$ nm and various waists $w_0 = 2, 2.5, 3, 4~\mu\text{m}$.

The coupling terms presented in this paper may hinder or benefit applications for levitated disks. Disks have been proposed as potential accelerometers for gravitational wave detection \cite{Arvanitaki2013}. The third order coupling terms may complicate determining which degree of freedom experienced a force or torque. On the other hand, it may be used as a means for indirectly detecting the motion of several degrees of freedom with a single detection scheme and therefore an efficient force/torque detector. Another common application is cooling the motion of the disk in attempt to study macroscopic quantum mechanics \cite{1367-2630-10-9-095020,Romero-Isart2011,Romero-Isart2011a}. As energy from one degree of freedom can be transferred to another through the couplings, it may have potential for sympathetically cooling several degrees of freedom by performing a cooling method on only one of the degrees of freedom. Preliminary results show that this is indeed possible for both radii studied using parametric feedback cooling or cold damping. Provided near ground state cooling can be achieved, the multi-mode resonator can further be used to explore entanglement or quantum state transfer \cite{PhysRevB.90.174307,PhysRevLett.109.013603}.


The coupling terms are found to scale as the square ratio of the radius to the beam waist, $a^2/w^2_0$, and may therefore have less of an impact on the dynamics for particles of smaller radii compared to the wavelength. It is also found that the influence of the coupling may be reduced by sufficiently separating each degree of freedom's harmonic frequency. 

This paper is organized as follows. Section \ref{analytical} illustrates the analytical calculation of the potential energy of thin dielectric disks in a Gaussian standing wave. The potential energy is approximated to reveal a term third order in displacements and rotations. In Sec. \ref{dda}, the procedure for calculating forces and torques on a disk using the discrete-dipole approximation is outlined. The corresponding coefficients/frequencies are presented for various Gaussian beam waists. Lastly, Sec. \ref{dynamics} examines the resulting dynamics due to the harmonic and coupling terms described in the previous sections. 

\begin{figure}[h]  
\centering
  \hspace*{-0.5cm}\includegraphics[width=0.45\textwidth]{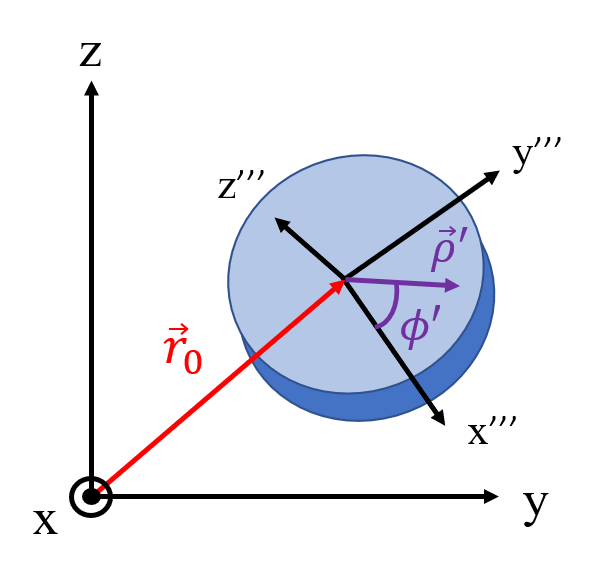}  
    \caption{Coordinate system in the lab frame $(x,y,z)$ and the particle frame $(x''',y''',z''')$. The disk's symmetry axis is aligned with the particle frame $z'''$ axis. The disk's center of mass as measured from the lab frame $\vec{r}_0$ is shown in red. The location to a point on the thin disk is given by the polar coordinates $(\rho^{\prime},\phi^{\prime})$ in the particle frame which are shown in purple. \label{fig:coordinates}}
\end{figure}
\section{Approximate Analytical Potential Energy\label{analytical}}
This section outlines the analytical calculation of the potential energy of a thin dielectric disk in a Gaussian standing wave in the Rayleigh-Gans approximation. In this approach the disk thickness is taken to be very thin so that the Rayleigh approximation holds along that direction \cite{doi:10.1063/1.326257,Stickler2016a}. The approximated results verify the existence and helps elucidate the origin of the terms responsible for the dynamics seen in the following sections. 

The disk is described with radius $a$, thickness $T\ll \lambda$, index of refraction $n$, and susceptibilities $\chi_{\parallel} = n^2-1$ and $\chi_{\perp}=\chi_{\parallel}/n^2$ corresponding to the susceptibility parallel and perpendicular to the disk symmetry axis ($z'''$ axis), respectively \cite{Stickler2016a}. The principal moments of inertia are $I_z = ma^2/2$ and $I_x=I_y=m(3a^2+T^2)/12$. The disk's center of mass is located at $\vec{r}_0=\langle x_0, y_0, z_0  \rangle$ and rotations are described in terms of the Euler angles ($\alpha ,\beta ,\gamma$) in the $z-y'-z''$ convention \cite{Seberson2019,SakuraiJ.J.JunJohn2011Mqm,MARION1965361}. 

The Gaussian standing wave is formed by two counter-propagating Gaussian waves with non-zero longitudinal components so that they satisfy Maxwell's equations \cite{novotny_hecht_2006}. Each traveling wave has the symmetric waist $w_0$, wavenumber $\vec{k}= (\pm \hat{x}) 2\pi/\lambda$, and is polarized in the $\hat{z}$ direction. Around the focus, $x=0$, each wave takes the form
\begin{equation}
\vec{E}_{\pm}(x,y,z) = E_0 e^{- \left(y^2 +z^2 \right)/w^2_0 }\left[ \hat{z} \mp \frac{iz}{x_R}\hat{x} \right]e^{\pm ikx}, \label{eq1} 
\end{equation}
where $x_R = kw^2_0/2$ is the Rayleigh range and $+$ ($-$) stands for the the right (left) traveling wave. The incident fields used for the numerical calculations in Secs. \ref{dda} and \ref{dynamics} are found by propagating Eq. \eqref{eq1} throughout all space using the angular spectrum representation \cite{novotny_hecht_2006}. For the analytical calculations performed in this section and in Appendix \ref{appendixB} the approximated Gaussian standing wave 
\begin{equation}
\vec{E}_{\rm inc}(x,y,z) \approx E_0 e^{- \left(y^2 +z^2 \right)/w^2_0 }\left[ \cos kx\hat{z} + \sin kx \frac{z}{x_R}\hat{x} \right], \label{eq2} 
\end{equation}
is used, which is valid in the space $|x|\ll x_R$. 

The mechanical potential energy associated with the interaction between the light and the dielectric is
\begin{equation}
U = -\frac{1}{4}\int \vec{P}(r')\cdot \vec{E}(r')  d^3r' , \label{eq3} 
\end{equation}
where the integral is over the volume of the disk, $\vec{P}(r') = \epsilon_0\overset{\text{\tiny$\leftrightarrow$}}{R}^\dagger \overset{\text{\tiny$\leftrightarrow$}}{\chi}_{0}  \overset{\text{\tiny$\leftrightarrow$}}{R} \vec{E}(r')$ is the polarization vector, $\overset{\text{\tiny$\leftrightarrow$}}{\chi}_{0}$ is the diagonal susceptibility matrix in the nanoparticle frame, and $\overset{\text{\tiny$\leftrightarrow$}}{R}$ is the rotation matrix. The rotation matrix in terms of the Euler angles explicitly can be found in Appendix \ref{appendixA}. The potential energy in the Rayleigh-Gans approximation with the incident field, Eq. \eqref{eq2}, becomes
\begin{align}
\begin{split}
U &\approx -\frac{E^2_0}{4}  \int e^{- 2\left[ y^2(r')  +z^2(r')  \right] / w^2_0 }   \\
& \qquad \times \left[ \cos^2 kx(r')\chi_1 + \sin 2kx(r') \frac{z(r') }{x_R} \chi_2 \right] d^3 r' ,  \label{eq4} 
\end{split}
\end{align}
where $\chi_1 = \Delta\chi\cos^2\beta +\chi_{\perp}$,  $\chi_2 = \Delta\chi\sin\beta\cos\beta\sin\alpha$, $\Delta\chi = \chi_{\parallel}-\chi_{\perp}$, and a higher order term proportional to $z^2(r')\sin^2kx(r')$ was dropped. To evaluate Eq. \eqref{eq4} the coordinates of the disk must be projected onto each lab frame coordinate ($x, y, z$) and it is favorable to move to polar coordinates. First, in the limit $T\ll \lambda$ the functions in Eq. \eqref{eq4} are independent of the thickness leaving the functions in the integral dependent only on the disk's radial and angular coordinates $(r') = (\rho', \phi ')$ (see Fig. \ref{fig:coordinates}). In terms of the center of mass and disk coordinates, $x(r'), y(r'),$ and $z(r')$ in Eq. \eqref{eq4} are 
\begin{equation}
x_i(r') = x_{0,i} + \pvec{\rho}' \cdot \hat{x}_i , \quad x_i = (x,y,z) , \label{eq5} 
\end{equation}
with
\begin{align}
\begin{split}
\pvec{\rho}' &= \overset{\text{\tiny$\leftrightarrow$}}{R}^\dagger \vec{\rho}   \\
&= \rho ' \left[  \cos\phi ' \begin{pmatrix} R_{11} \\ R_{12} \\ R_{13} \end{pmatrix} +\sin\phi '  \begin{pmatrix} R_{21} \\ R_{22} \\ R_{23} \end{pmatrix} \right] , \\ \label{eq6} 
\end{split}
\end{align}
and the $R_{ij}$ are matrix components in the rotation matrix $\overset{\text{\tiny$\leftrightarrow$}}{R}$ (see Appendix \ref{appendixA}). Insertion of Eqs. \eqref{eq5} and \eqref{eq6} into Eq. \eqref{eq4} leads to analytic solutions in terms of Bessel functions. In the limit of small radius $w_0 \gg a$, $r_0a\ll w^2_0$ where the zeroth order approximation to the exponentials ($\sim 1$) can be used, a Bessel function of the first kind is obtained as was found in Ref. \cite{Stickler2016a}. However, this approximation misses the coupling of the disk to the Gaussian standing wave and a fourth order expansion in the coordinates is required to resolve it. 

Practical parameters in levitated optomechanics are in the range $(\lambda, w_0) \sim 1~\mu\text{m}$ and $(r_0, a) \sim 1-0.1~\mu\text{m}$. For the derivation, the limits $a^2\ll w^2_0$, $r^2_0\ll w^2_0$ are used to expand each function in Eq. \eqref{eq4} to fourth order in the coordinates and terms $\mathcal{O}(a^6/w^6_0)$ as well as $\mathcal{O}(x^n_{0,i}\pi^m_j)$, where $n+m\geq 4$, $\pi_j=(\alpha, \beta)$, are dropped which retains terms up to third order in the coordinates. Due to the symmetry of the disk, the potential energy is independent of the angle $\gamma$. Further, the disk's symmetry axis is primarily aligned along the lab frame $\hat{x}$ direction and, as will be justified in the next section, rotates at angles that justify the small angle approximation $\alpha \rightarrow 0+\theta_z$ , $\beta \rightarrow \pi/2 +\theta_y$ with $\theta_z$, $\theta_y$, small. Here $\theta_z$ represents small angle rotations about the lab frame $z$ axis while $\theta_y$ is a small rotation about the lab frame $y$ axis. The resulting potential energy is of the form
\begin{align}
\begin{split}
U &\approx \frac{m}{2} \left( \omega^2_x x^2_0 + \omega^2_y y^2_0 + \omega^2_z z^2_0  \right)  +  \frac{I_x}{2} \left( \omega^2_{\theta_y} \theta^2_y + \omega^2_{\theta_z} \theta^2_z  \right)   \\
& \qquad \qquad + m x_0\left( \omega^2_1 y_0 \theta_z - \omega^2_2  z_0 \theta_y  \right).  \label{eq7}
\end{split}
\end{align}
Explicit expressions for the $\omega_i$ may be found in Appendix \ref{appendixB}. The terms in the first row in the above potential describe simple harmonic motion for the three translational and two rotational degrees of freedom. The last term is a coupling between the translational and rotational degrees of freedom that is of third order in the coordinates. The coupling terms arise due to the finite radius of the disk and the Gaussian and standing wave geometry of the beam. An asymmetric electric field gradient across the disk produces a stronger force on the section of the disk with greater laser intensity. That section of the disk is pulled into the region of the trap with greater laser intensity more strongly than the section of the disk with less field intensity. As the radius increases and the trap becomes more confining, the greater the electric field gradient across the disk and the more influential the coupling terms are. With reference to Eq. \eqref{eq4}, it is a result of the ro-translational coupling in the Gaussian together with the $x_0$ dependence in $\cos^2 k x(r')$ describing the standing wave. The asymmetry in the $(\omega_1,\omega_2)$ coefficients is due to the $\hat{x}$ component of the incident electric field proportional to $z_0/x_R$. If this term is negligible, $x_R\gg z_0$, the coefficients are equivalent, $\omega_1 = \omega_2$. 

To garner an idea of the dynamics that arise due to the coupling, consider the $x_0y_0\theta_z$ term in Eq. \eqref{eq7}. A disk displaced by $\vec{r}_0 = \langle x_0, y_0, 0  \rangle$ in Fig. \ref{fig:coordinates} experiences a torque about the $-z$ axis due to a greater electric field intensity on the side of the disk nearest the focus. These terms are therefore a gradient force/torque as a consequence of the electric field gradient along the finite extension of the disk. 
\section{Numerical Evaluation of the Forces and Torques \label{dda}}
\subsection{System and Procedure \label{system}}
The optical scattering problem for finite sized dielectric objects is generally difficult to solve analytically. As was done in the previous section, approximations are often required to glean insight into the dynamics. Another rigorous approach is to numerically solve for the scattered electromagnetic waves and use the resulting Maxwell stress tensor to obtain the forces and torques. This section details the results from performing the latter method by numerically implementing the discrete-dipole approximation (DDA) to calculate the scattered fields of the disk \cite{1988ApJ...333..848D,Draine:94}. 

In the DDA, the disk is composed of $N$ discrete spherical dipoles each with polarizibility $\alpha$ and the internal fields of the dielectric are solved for self-consistently to retrieve the scattered fields outside the particle. In the implementation of the DDA used for this paper, each dipole that composed the spherical dipole had a polarizibility $\alpha = 4\pi\epsilon_{0}R^{3}\left( \dfrac{n^2-1}{n^2+2} \right)$. The method developed has been shown to be accurate to within 1\% by comparing the scattered fields from a discretized sphere to the exact Mie scattering solutions \cite{BohrenCraigF.2004Aaso}. The scattered fields that are generated from the DDA are then added to the incident field and inserted into the Maxwell stress tensor \cite{jackson_classical_1999}
\begin{equation}
T_{ij} = \epsilon_0 \left[ E_iE_j + c^2B_iB_j -\frac{1}{2} \left( |\vec{E}|^2 + c^2|\vec{B}|^2  \right)\delta_{ij} \right], \label{eq8}
\end{equation}
in order to obtain the forces and torques 
\begin{align}
\vec{F} &= \oint  \overset{\text{\tiny$\leftrightarrow$}}{T} \cdot \hat{n} \,dS , \label{eq9} \\
\vec{\tau} &= \oint \overset{\text{\tiny$\leftrightarrow$}}{M}   \cdot \hat{n}      \,dS , \label{eq10}
\end{align}
where $\overset{\text{\tiny$\leftrightarrow$}}{M} = -\overset{\text{\tiny$\leftrightarrow$}}{T}\times \vec{r} $. The surface over which the integration is performed was taken to be a sphere centered at the disk center with radius $1.5\times$ that of the disk. The surface integration was performed using Gaussian quadrature with increasing number of points until convergence was demonstrated.

The above procedure was performed for dielectric disks located near the intensity maximum of a Gaussian standing wave. To construct the standing wave, a right-traveling wave, $\vec{E}_{R}(x, y, z)$ is found by propagating Eq. \eqref{eq1} throughout all space using the angular spectrum representation with no paraxial approximation. A left-traveling wave, $\vec{E}_{L}(x, y, z) = \vec{E}_{R}(-x, -y, z)$, is added to the right-traveling wave to form the standing wave. The wavelength of each wave is $\lambda = 850\,\rm nm$  and is fixed throughout this paper. While the detailed coefficients of the forces and torques change with wavelength the major results of this paper do not, and $850\,\rm nm$  is an efficient emission wavelength for GaAs quantum well gain media used in semiconductor lasers \cite{grine_etal}. A range of Gaussian beam waists were explored $w_0 = 2, 2.5, 3, 4\,\mu\text{m}$ to define the optical trap. 

Most of the calculations performed were for disks of radius $a=2~\mu\text{m}$, thickness $T=\lambda/4n$ to achieve maximum light coupling, and index of refraction $n=1.45$ or $n=2.0$. The indices of refraction correspond to materials composed of silica and silicon nitride, respectively. Unless otherwise stated, the data and discussions that follow will refer to this set of parameters. 

The following example outlines the steps for how a calculation is performed: the disk's symmetry axis is aligned with the axial direction ($x$ axis), the disk is displaced a distance $y_0$ from the focus of the standing wave, the scattered waves are calculated using the DDA, the forces and torques are computed using Eqs. \eqref{eq9} and \eqref{eq10}. The process is identical for rotations: the disk is initially situated at $\vec{r}_0= \langle 0, 0, 0 \rangle$ and $(\alpha=0, \beta=\pi/2)$, a rotation is made $\alpha=0+\theta_z$, the scattered waves are calculated using the DDA, the forces and torques are calculated.  The baseline for the calculations is when the disk is placed symmetrically at the focus of the standing wave, $\vec{r}_0= \langle 0, 0, 0 \rangle$ , $(\alpha=0, \beta=\pi/2)$ which should be a potential minimum. Indeed, a force or torque due to a displacement generally gives a value at least ten orders of magnitude greater than the baseline. 

\subsection{Forces and Torques \label{coefficients}}
As is expected in levitated optomechanics, small displacements in one direction reveals a spring force in that same direction $F_i = -k_ix_{i,0}$ and torque $\tau_i = -\kappa_i\pi_i$, $\pi_i = (\alpha, \beta)$. The spring constants for each degree of freedom, ($k_i, \kappa_i$), are determined by direct division, $k_i = -F_i/x_{i,0}$. At the harmonic level, no coupling of the different degrees of freedom through the potential energy were found.  

Being that there are 6 degrees of freedom (including $\gamma$), there are 15 different second order couplings possible in the forces and torques. Of these possibilities, only terms similar to that in Eq. \eqref{eq7} were found to be above the baseline. These terms were found to be significant for disks of large and small radii. For example, a displacement of the center of mass by $\vec{r}_0 = \langle x_0, 0 ,z_0  \rangle$ produces a torque about the $y$ axis, suggesting a term in the potential energy $U\propto D_1y_0z_0\theta_y$, with $D_1$ a proportionality constant. A similar coupling of the same order was found $U\propto D_2x_0y_0\theta_z$, with $D_2\neq D_1$ necessarily. The coefficients $D_1$ and $D_2$ are also determined by division, i.e. $D_1 = F_z/(y_0\theta_x)$. Interestingly, the coefficients computed in this way generally gives different values for the force in the $\hat{y}$ and $\hat{z}$ directions
\begin{align}
F_y &\propto -Ax_0\theta_z, \label{eq11} \\
F_x &\propto  -C_1y_0\theta_z , \label{eq12} \\
\tau_z &\propto  -C_1x_0y_0, \label{eq13} 
\end{align}
for the first coupling term, and
\begin{align}
F_z &\propto Bx_0\theta_y, \label{eq14} \\
F_x &\propto C_2z_0\theta_y , \label{eq15} \\
\tau_y &\propto  C_2x_0z_0, \label{eq16} 
\end{align}
for the second coupling term, with $A\approx C_1$ and $B\approx C_2$. The coefficients $A$ and $B$ can differ from $C_1$ and $C_2$ by 2\% using a waist of $w_0= 2~\mu\text{m}$ and 20\% using a waist of $w_0= 4\,\mu\text{m}$. Although the discrepancy is suspected to be due to higher order terms, we are only interested in the dynamics due to this term and the average values $D_1 = (A+2C_1)/3$ and $D_2 = (B+2C_2)/3$ will be used from here on so that potential energy can be written in the form of Eq. \eqref{eq7}. The consequences of using the average values is insignificant and will be discussed in Sec. \ref{dynamics}.

\begin{figure}[h]  
\centering
  \hspace*{-0.cm}\includegraphics[width=0.48\textwidth]{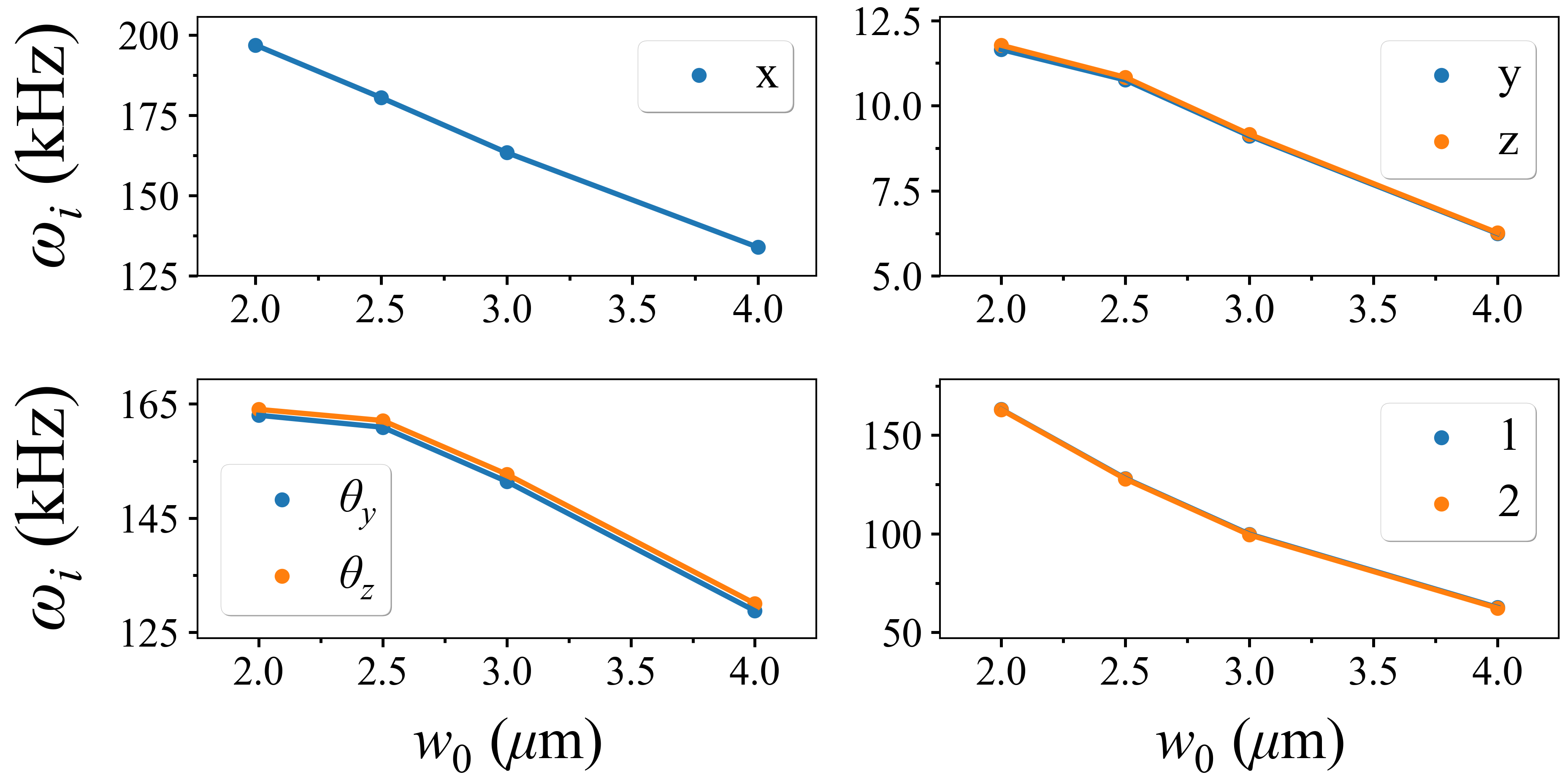}  
    \caption{Frequencies obtained using the DDA for silica disks (n=1.45) of radius $a= 2~\mu\text{m}$ and thickness $T=\lambda/(4n)$ for varying beam waist. A fixed total power of 100 mW is used for the calculations.  For each calculation the disk was composed of $N=299744$ points with a thickness of 8 points.\label{fig:waist} }
\end{figure}

\begin{figure}[h]  
\centering
  \hspace*{-0.cm}\includegraphics[width=0.48\textwidth]{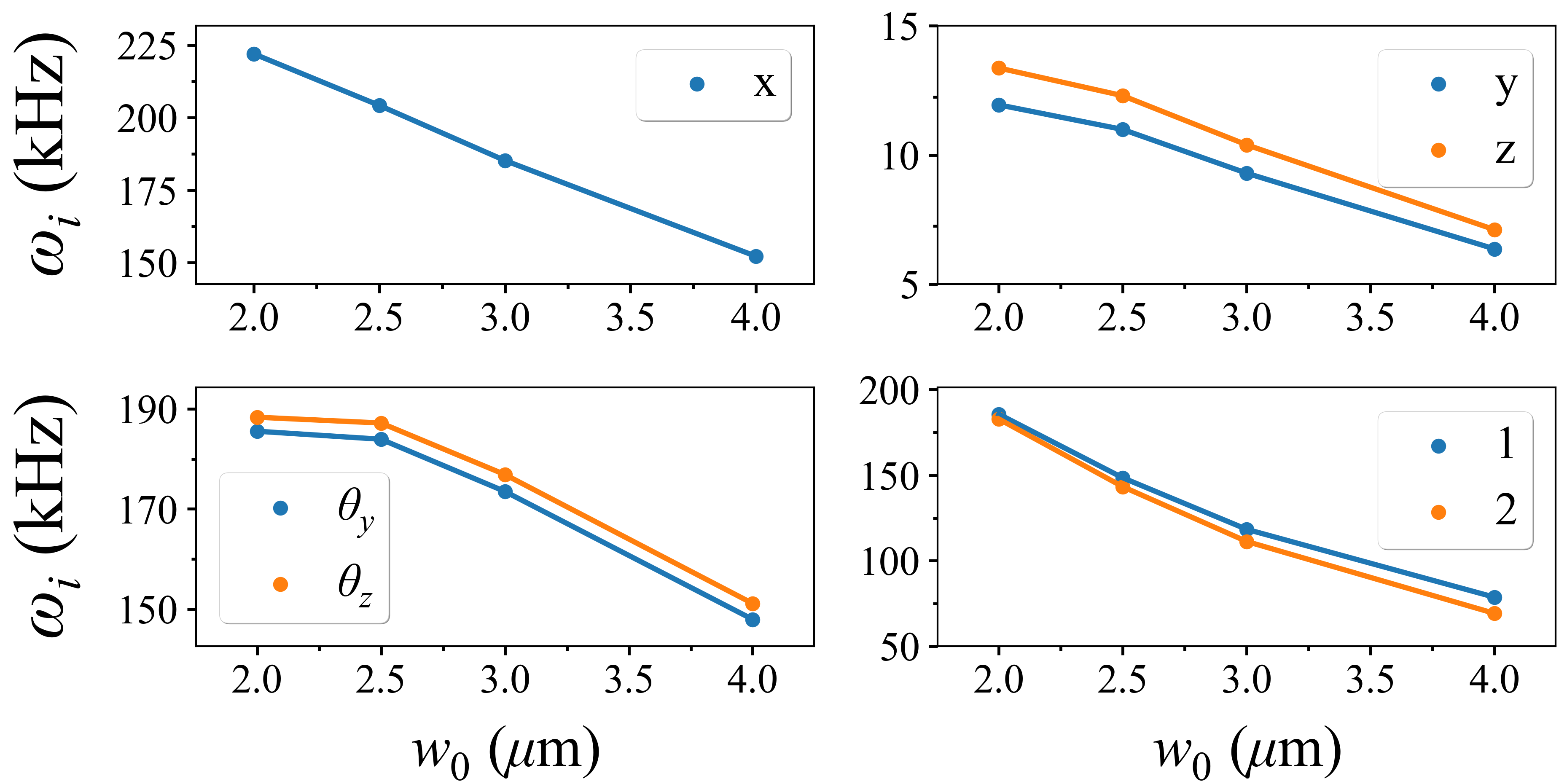}  
    \caption{Frequencies obtained using the DDA for silicon nitride disks (n=2.0) of radius $a= 2~\mu\text{m}$ and thickness $T=\lambda/(4n)$ for varying beam waist. A fixed total power of 100 mW is used for the calculations.  For each calculation the disk was composed of $N=569984$ points with a thickness of 8 points. \label{fig:waist2} }
\end{figure}

 \begin{table*}[t]
 \centering
 \begin{ruledtabular}
 \begin{tabular}{c | c | c | c | c | c | c | c} %
 $w_0$ ($\mu$m) & $\omega_x$ (kHz) & $\omega_y$ (kHz)  &  $\omega_z$ (kHz) & $\omega_{\theta_y}$ (kHz) & $\omega_{\theta_z}$ (kHz) & $\omega_1$ (kHz) & $\omega_2$ (kHz)\\
  \hline
 2 & 394 & 38 & 38 & 537 & 390 & 46 & 39\\
    \hline
 3 & 264 & 17 & 17 & 361 & 263 & 21 & 17\\
 \end{tabular}
 \end{ruledtabular}
  \caption{\label{table1} Frequencies for a silica disks of radius $a= 200$ nm and thickness $T=\lambda/(40n)$ for two beam waists $w_0 = 2 , 3~\mu\text{m}$. The disk has dimensions that are reduced by a factor of ten from the $a = 2~\mu\text{m}$, $T=\lambda/(4n)$ disks. A fixed total power of 100 mW is used for the calculations. The number of points used to compose the disk was $N= 37488$ and the thickness of the disk was 4 points.  }
 \end{table*}

The spring and coupling constants $(k_i, \kappa_i, D_i)$ have the same units and are most useful when written in terms of frequencies
\begin{equation}
\omega_i = \sqrt{k_i/m} ,\quad i=(x,y,z),  \label{eq17}
\end{equation}
for translational harmonic motion,
\begin{equation}
\omega_i = \sqrt{\kappa_i/I_x}  ,\quad i=(\theta_y,\theta_z),  \label{eq18a}
\end{equation}
for rotational harmonic motion, and
\begin{equation}
\omega_i = \sqrt{D_i/m}  ,\quad i=(1,2),  \label{eq19a}
\end{equation}
for the coupling terms. 

Values for the frequencies as a function of beam waist are shown in Fig. \ref{fig:waist} for silica and Fig. \ref{fig:waist2} for silicon nitride using a fixed total laser power of 100 mW. The general trend identified from the figures is that each frequency decreases as the waist increases. This feature is not unexpected, however, for particles in the Rayleigh regime $\lambda\gg a$, $\omega_i\propto 1/w^2_0$ while for $a=2~\mu\text{m}$ disks the dependence is nearly linear. 

For both materials, the frequency in the axial direction is in the $150-200$ kHz range while the radial degrees of freedom oscillate in the $1-10$ kHz range. The axial frequency is most strongly affected by the standing wave which is independent of the waist. However, the radial frequencies are dominantly due to the Gaussian geometry. To leading order (see Appendix \ref{appendixB}), for fixed power the axial frequencies depend inversely on the wavelength and waist $\omega_x \propto 1/\lambda w_0$ while the radial frequencies depend on the waist as $\omega_{y,z} \propto 1/w^2_0$, hence the disparity between the axial and radial frequencies. Note that part of the waist dependence on each frequency is due to the dependence of the laser intensity on the waist. Each frequency therefore shares a $1/ w_0$ dependence from the power. 

For a $2~\mu\text{m}$ radius disk at $T=300$ K, these frequencies correspond to translational oscillation amplitudes of $x_0\sim 1$ nm and $(0,y_0, z_0)\sim 20$ nm. The rotational frequencies are closer to the axial frequency and in the range $190-125$ kHz. The rotational frequencies differ by 20\% between the two materials at the same waist. Using the average frequency, this corresponds to angular displacements of $\sim 1$ mrad. Displacements of this size justify some of the approximations made in Sec. \ref{analytical} since $r_0\ll w_0$ and $\sin\alpha \approx \theta_z$. 

Also shown in Figs. \ref{fig:waist} and \ref{fig:waist2} are the coupling coefficients $(\omega_1, \omega_2)$. The coefficients being in the $50-200$ kHz range are comparable to both the rotational and axial frequencies. Due to the large coupling frequencies combined with the relatively large oscillation amplitude in the radial degrees of freedom, the resulting forces/torques due to the coupling terms have an impact on the dynamics as shown in Sec. \ref{dynamics}. 

Force and torque calculations were also performed for silica disks of radius $a= 200$ nm and thickness $T=\lambda/(40n)$ for the two beam waists $w_0 = 2 , 3~\mu\text{m}$. The dimensions are $10\times$ smaller than the $a = 2~\mu\text{m}$, $T=\lambda/(4n)$ disk. The resulting frequencies are shown in Table \ref{table1}. From the table, each frequency scales as $\omega_i\sim 1/w_0$ except for the radial frequencies $(\omega_y,\omega_z)\sim 1/w^2_0$. This dependence on the waist is consistent with the analytical frequencies given in Appendix \ref{appendixB}. Also from the table, each harmonic frequency is larger, and the coupling frequencies reduced, compared to its $a= 2~\mu\text{m}$ and $T=\lambda/(4n)$ counterpart in Fig. \ref{fig:waist}. The dependence of each frequency on the radius is also consistent with that found analytically in Appendix \ref{appendixB}. The harmonic frequencies increase as the radius decreases since the disk has greater field intensity per volume. The coupling frequencies scale as $\sim a/w^2_0$ due to the electric field gradient across the disk. This dependence provides a factor of ten between the $a=200$ nm and $a = 2~\mu\text{m}$ coupling frequencies. 

\subsection{Accuracy of the DDA\label{accuracy}}

\begin{figure}[h]  
\centering
  \hspace*{-0.cm}\includegraphics[width=0.48\textwidth]{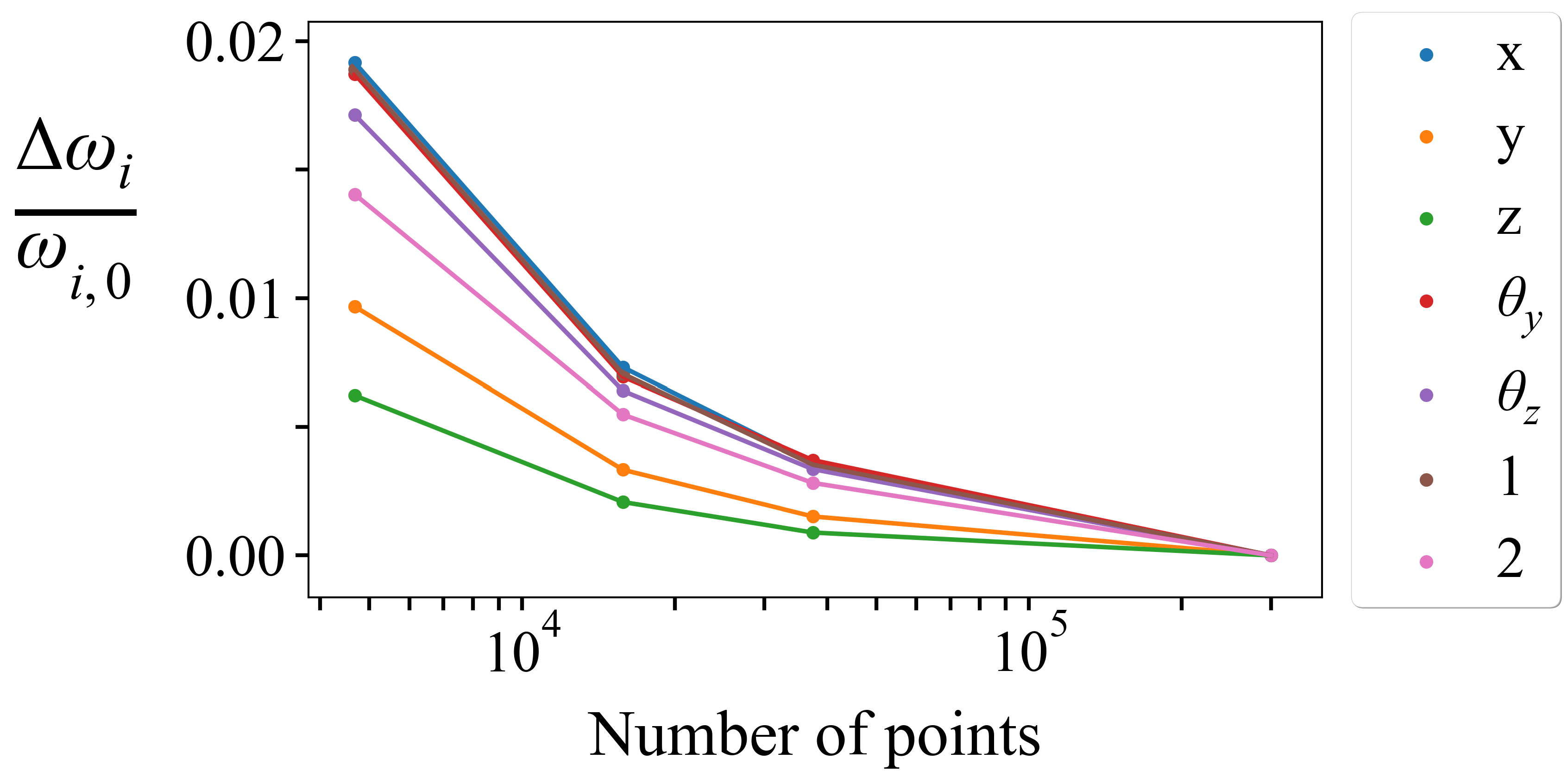} 
    \caption{Frequencies obtained for a $a=2~\mu\text{m}$ silica disk using DDA for varying number of points that the disk was composed of relative to the frequency obtained using 299744 points, $\omega_{i,0}$. The legend describes the various frequencies for the $x, y, z, \theta_y, \theta_z$ degrees of freedom as well as the $\omega_1, \omega_2$ coupling frequencies. The data points along the $x$-axis are 4680, 15804, 37488, and 299744 points. Comparing the left and rightmost data points in the figure shows that using 64 times more points changes the frequencies by less than 2\%.  \label{fig:points}}
\end{figure}

The frequencies shown in Sec. \ref{coefficients} were obtained through several numerical operations such as integrations and the implementation of the DDA. One of the major questions regarding convergence of these values is how many points (i.e. number of discrete dipoles), $N$, should be used to discretize the disk. Figure \ref{fig:points} shows the relative change of the various frequencies discussed in the previous subsections as a function of the number of points used to compose the disk. Here, $\omega_{i,0}$ is the frequency calculated using the largest number of points shown in the plot, $N=299744$. The frequency calculated using $N$ points is $\omega_i$. The change in the frequency $\omega_i$ compared to $\omega_{i,0}$ points is then $\Delta\omega_i = \omega_{i}-\omega_{i,0}$. The plot is shown for all of the various frequencies discussed above using a $a=2~\mu\text{m}$ silica disk with a $w_0=2~\mu\text{m}$ waist. Increasing the number of points by a factor of 64 from $N=4680$ to $N=299744$ changes the frequency by less than 2\%. On the other hand, the time complexity of the DDA method used to calculate the scattered light from the disk scales as $N\ln{N}$.
\section{dynamics\label{dynamics}}
The previous two sections have illustrated that disks levitated in Gaussian standing waves experience simple harmonic motion as well as non-harmonic forces and torques involving second order couplings. This section discusses the resulting dynamics due to these forces and torques as well as the natural torques that arise in rigid body dynamics. 

Thus far the focus has been on identifying terms in the potential energy. For translational motion the kinetic energy is trivial and leads to the equations of motion 
\begin{align}
\ddot{x} &= -\omega^2_x x - \left( \omega^2_1y\theta_z  - \omega^2_2z\theta_y \right), \label{eq18} \\ 
\ddot{y} &= -\omega^2_y y - \omega^2_1x\theta_z , \label{eq19} \\ 
\ddot{z} &= -\omega^2_z z + \omega^2_2x\theta_y , \label{eq20} 
\end{align}
for small angle oscillations. 

As was shown in Ref. \cite{Seberson2019}, for a symmetric top-like rigid body the rotational kinetic energy naturally involves coupling between the $\alpha, \dot{\alpha}, \beta$, and $\dot{\beta}$ degrees of freedom. Whether these terms are significant or not depends on the geometry. For $a=200$ nm disks each non-linear coupling term is significant and must be considered. For $a=2~\mu\text{m}$ disks, the term responsible for precession about the $x$ axis is the largest, but is still $10^{-4}$ times smaller than the harmonic term and is therefore negligible. The equations of motion for $a=2~\mu\text{m}$ disks are then written as
\begin{align}
\ddot{\theta}_y &= -\omega^2_{\theta_y} \theta_y + \frac{m}{I_x}\omega^2_2xz , \label{eq21} \\    
\ddot{\theta}_z &= -\omega^2_{\theta_z} \theta_z - \frac{m}{I_x}\omega^2_1xy , \label{eq22} \\     
\dot{\gamma} &= \omega_3 = const, \label{eq23} 
\end{align}
for small angle oscillations. 

\begin{figure}[h]  
\centering
  \hspace*{-0.cm}\includegraphics[width=0.48\textwidth]{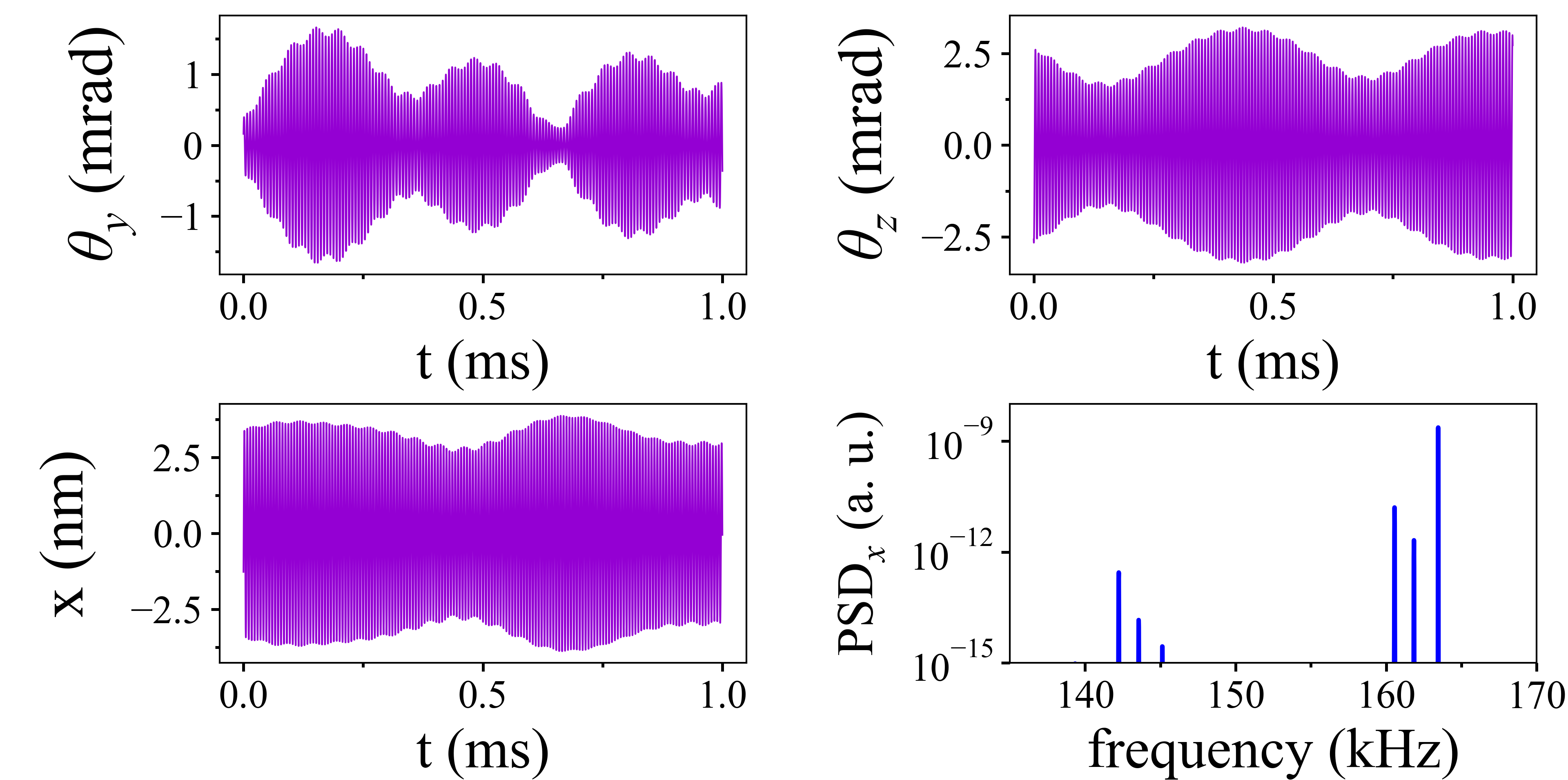} 
    \caption{Example trajectories of the $x$ and two rotational degrees of freedom as well as the power spectral density of the axial motion for a $a=2~\mu\text{m}$ silica disk in a $w_0=3~\mu\text{m}$ waist trap. The influence of the second order coupling term produces several amplitude modulations at different frequencies, but the disk remains stable. The frequencies of modulation in the $x$ degree of freedom can be seen in the power spectral density. Note that the rotational amplitudes remain in the $\sim$ mrad range, justifying the small angle approximation.  \label{fig:trajectories}}
\end{figure}

Figure \ref{fig:trajectories} shows sample trajectories of the $\theta_y$, $\theta_z$, and $x$ motions of a $a=2~\mu\text{m}$ silica disk in a $w_0=3~\mu\text{m}$ waist Gaussian standing wave by simulating Eqs. \eqref{eq18} to \eqref{eq22} at $T=300$ K. The influence of the second order coupling terms are seen to be significant for the three degrees of freedom with each trajectory containing modulations at various frequencies. Without the couplings the oscillations would be at the same amplitude for all times. In a gaseous environment these modulations might be mistaken for noise in an experiment. 

The bottom-rightmost plot in Fig. \ref{fig:trajectories} shows the power spectral density (PSD) of the $x$ motion. The harmonic frequency $\omega_x/2\pi = 163$ kHz is the largest and rightmost peak in the PSD. The other frequencies in the figure are the harmonic frequency plus the sums and differences of the various $y, z, \theta_y,$ and $\theta_z$ frequencies. Whereas sidebands due to coupling typically appear symmetrically on each side of the harmonic frequency, the frequency structure seen in Fig. \ref{fig:trajectories} is such that all significant modes have smaller frequency than the harmonic frequency. This is not a general feature of the coupling term and depends on the degree of freedom that is being observed and the various levels of degeneracy. 

In their analysis of disks as gravitational wave detectors, Ref. \cite{Arvanitaki2013} considered plane waves to form the standing wave in which case the couplings would be absent (see Sec. \ref{analytical}). A single peak in the PSD can then be used to describe a translational degree of freedom. However, a Gaussian standing wave is needed to trap the non-axial degrees of freedom and therefore should be considered. A concern then for the disks ability to be a detector is the extra 'noise' the degree of freedom being observed for detection of the wave will have. If there is a disturbance in one of the degrees of freedom, the degree of freedom being observed for gravitational waves will also be disturbed to some degree that depends on the coupling.

The influence of the coupling term on each degree of freedom has two factors: the size of the coefficients $\omega_1$ and $\omega_2$, and the level of degeneracy of the coupled degrees of freedom. First, the $\omega_1$ and $\omega_2$ coupling coefficients are relatively large $\sim 100\, \rm kHz$. Second, strong coupling is achieved when the frequencies are nearly degenerate. Because $\omega_x$, $\omega_{\theta_y}$, and $\omega_{\theta_z}$ are close in frequency the coupling term produces a larger effect on these degrees of freedom. Since the radial degrees of freedom oscillate $10\times$ slower, the influence of the coupling term is significantly reduced, but not absent. 

The question of stability is one of the most important for applications using levitated nanodisks. Despite the seemingly complicated motion, simulations have shown no evidence that this motion is unstable. The disk remains stable in the trap after several thousand oscillations for all of the beam waists explored $w_0 = 2, 2.5, 3, 4~\mu\text{m}$. The $a=200$ nm disk was found to be stable at all frequencies, even with inclusion of the non-linear coupling terms in the rotational kinetic energy \cite{Seberson2019}. Recall from Sec. \ref{coefficients} the differing coefficients in Eqs. \eqref{eq11} to \eqref{eq16} as produced from the DDA calculations. Simulating the equations of motion with different coefficients attached to each degree of freedom's coupling term causes no issue for stability. 

A common application in levitated optomechanics is cooling the motion of the levitated particle in attempt to reach the ground state, or to reach lower pressures \cite{PhysRevA.101.053835}. The couplings in this paper offer a possibility of cooling one or more degrees of freedom sympathetically by actively cooling only one degree of freedom. The full dynamics of cooling using the couplings is beyond the scope of this paper, but we note some preliminary findings. Through simulations of the equations of motion Eqs. \eqref{eq18} to \eqref{eq22}, results show that sympathetic cooling is indeed possible. For both radii, parametric feedback or cold damping \cite{Gieseler2012,Li2011} is an effective method for cooling multiple degrees of freedom. By inserting artificial numbers for the frequencies in the simulation, two relations were found for optimal cooling. Frequencies tailored within a few kHz of the relations $\omega_x = \omega_{\theta_y} \pm \omega_z$ and/or $\omega_x = \omega_{\theta_z} \pm \omega_y$, can achieve significant sympathetic cooling to at least the mK regime. From Fig. \ref{fig:waist}, $a=2~\mu\text{m}$ silica disks are naturally in this regime. From Appendix \ref{appendixB}, each frequency depends on several parameters and has the possibility to be tuned to achieve optimal cooling experimentally.  
\section{Conclusion}
The forces and torques exerted on dielectric disks trapped in a Gaussian standing wave were analyzed for disks of radius $2~\mu\text{m}$ with index of refraction $n=1.45$ and $n=2.0$ as well as disks of radius 200 nm with $n=1.45$. Calculations of the forces and torques were conducted both analytically and under numerical simulation using a discrete-dipole approximation method. 

Similar to nanodumbbells, a nanodisk experiences restoring forces in all three translational degrees of freedom, restoring torques in two rotational degrees of freedom, and has constant spin about the symmetry axis. Due to the finite geometry of the disk, third order, ro-translational coupling terms in the potential energy are found to be a necessary consideration when describing the dynamics of disks. The coupling terms are the result of an electric field gradient across the disk and depend on the ratio of the radius to the beam waist and on the temperature. 

The ro-translational coupling produces several modes of oscillation in the coupled degrees of freedom which are evident in the power spectral density. While the restoring forces are dominant, the coupling terms can become sizable through strong coupling, which manifests when the coupled degrees of freedom are nearly degenerate. Despite the couplings, simulations show no evidence that the motion is unstable, which is of utmost importance for applications such as gravitational wave detection, force sensing, and ground state cooling. 

\begin{acknowledgments}
Supported by the Laboratory Directed Research and Development program at Sandia National Laboratories, a multimission laboratory managed and operated by National Technology and Engineering Solutions of Sandia LLC, a wholly owned subsidiary of Honeywell International Inc. for the U.S. Department of Energy's National Nuclear Security Administration under contract DE-NA0003525. This paper describes objective technical results and analysis. Any subjective views or opinions that might be expressed in the paper do no necessarily represent the views of the U.S. Department of Energy or the United States Government. 

We would like to acknowledge Alejandro Grine, Darwin Serkland, Justin Schultz, Michael Wood, Peter Schwindt, and Tongcang Li for motivation of pursuit on the topic and useful discussions. 

\end{acknowledgments}

\appendix
\section{Rotation matrix \label{appendixA}}
The rotation matrix in the $z-y'-z''$ convention is 
\begin{align}
\overset{\text{\tiny$\leftrightarrow$}}{R} &=  \begin{pmatrix}  \text{c}\beta\text{c}\alpha\text{c}\gamma-\text{s}\alpha\text{s}\gamma & \text{c}\beta\text{s}\alpha\text{c}\gamma+\text{c}\alpha\text{s}\gamma & -\text{s}\beta\text{c}\gamma \\
			-\text{c}\beta\text{c}\alpha\text{s}\gamma-\text{s}\alpha\text{c}\gamma  & -\text{c}\beta\text{s}\alpha\text{s}\gamma+\text{c}\alpha\text{c}\gamma & \text{s}\beta\text{s}\gamma \\
																									\text{s}\beta\text{c}\alpha & \text{s}\beta\text{s}\alpha & \text{c}\beta \end{pmatrix}  \\
	&= \begin{pmatrix}
	R_{11} & R_{12} & R_{13} \\
	R_{21} & R_{22} & R_{23} \\
	R_{31} & R_{32} & R_{33} 
	\end{pmatrix}, \label{eqA1}
\end{align}
where the notation $\text{c}=\cos$, $\text{s}=\sin$ was used. 

For a disk with negligible thickness, a point on the disk is located at $\vec{\rho} = \rho \left( \cos\phi \hat{x} + \sin\phi\hat{y}  \right)$ in the body frame. In the lab frame, the point is located at $\pvec{\rho}' = \overset{\text{\tiny$\leftrightarrow$}}{R}^\dagger \vec{\rho} \equiv \langle \rho^{\prime}_x, \rho^{\prime}_y , \rho^{\prime}_z  \rangle$ yielding Eq. \eqref{eq6} in Sec. \ref{analytical}.

\section{Approximate analytical frequencies of motion \label{appendixB}}
The potential energy in Eq. \eqref{eq4} has the two assumptions $x_0\ll x_R$ and $T\ll \lambda$. In order to obtain a potential energy of the form Eq. \eqref{eq7} we further require a radius small compared to the waist and small displacements relative to the waist  $a^2\ll w^2_0$, $r^2_0\ll w^2_0$. As discussed in Sec. \ref{coefficients} the limit set upon the displacements are justified. Expanding each function to first order gives
\begin{align}
\begin{split}
U &= -\frac{TE^2_0}{4} \left[1- \frac{2\rho^2_0}{w^2_0} \right] \int^a_0 \rho^{\prime} d\rho^{\prime} \int^{2\pi}_0 d\phi^{\prime} \\
& \quad \times \left[1- \frac{2}{w^2_0}\left( \rho^{2\prime}_{y}  + \rho^{2\prime}_z \right)  \right] \\
& \quad \times \left[1- \frac{4}{w^2_0}\left( y_0\rho^{\prime}_y +  z_0\rho^{\prime}_z \right)  \right] \\
& \quad \times \left[ \chi_1\left( 1-\frac{k^2}{2} (x_0 + \rho^{\prime}_x)^2 \right)^2  \right. \\
\qquad &+  \chi_2\frac{2k}{x_R} \left( z_0 + \rho^{\prime}_z \right)\left( x_0 + \rho^{\prime}_x \right) \\ 
\qquad &+ \left. \chi_{12}\frac{k^2}{x^2_R} \left( z_0 + \rho^{\prime}_z \right)^2\left( x_0 + \rho^{\prime}_x \right)^2   \right],  \label{eqA2} 
\end{split}
\end{align}
where $\rho^2_0 = y^2_0+z^2_0$, the $\rho^{\prime}_i$ are defined in Appendix \ref{appendixA} above, $\chi_1 = \Delta\chi\cos^2\beta +\chi_{\perp}$,  $\chi_2 = \Delta\chi\sin\beta\cos\beta\sin\alpha$, $\chi_{12} = \Delta\chi\sin^2\beta\sin^2\alpha +\chi_{\perp}$, $\Delta\chi = \chi_{\parallel}-\chi_{\perp}$, and the integral over the thickness was performed. After carrying out the multiplications and integrations, terms $\mathcal{O}(a^6/w^6_0)$ as well as $\mathcal{O}(x^n_{0,i}\pi^m_j)$, where $n+m\geq 4$, $\pi_j=(\alpha, \beta)$, are dropped. Odd powers of $\rho^{\prime}_i\rho^{\prime}_j$ integrate to zero from the $\phi^{\prime}$ dependence. Lastly, from Sec. \ref{coefficients}, the disk rotates at angles that justify the small angle approximation $\alpha \rightarrow 0+\theta_z$ , $\beta \rightarrow \pi/2 +\theta_y$. The third order coupling terms originate from rows 3 and 4 in Eq. \eqref{eqA2} which are part of the Gaussian and standing wave geometries, respectively, 
\begin{align}
& \left( y_0\rho^{\prime}_y +  z_0\rho^{\prime}_z  \right)  \left(x_0 + \rho^{\prime}_x \right)^2 \\
& \propto \left( y_0\rho^{\prime}_y +  z_0\rho^{\prime}_z  \right)\left(x_0 \rho^{\prime}_x \right) \\
&\propto -x_0 \left( y_0\theta_z - z_0\theta_y \right). \label{eqA2a}
\end{align}
The resulting potential energy is of the form of Eq. \eqref{eq7}
\begin{align}
\begin{split}
U &\approx \frac{m}{2} \left( \omega^2_x x^2_0 + \omega^2_y y^2_0 + \omega^2_z z^2_0  \right)  +  \frac{I_x}{2} \left( \omega^2_{\theta_y} \theta^2_y + \omega^2_{\theta_z} \theta^2_z  \right)   \\
& \qquad \qquad + m x_0\left( \omega^2_1 y_0 \theta_z - \omega^2_2  z_0 \theta_y  \right),  \label{eqA3a}
\end{split}
\end{align}
and the analytical frequencies are 
\begin{align}
\omega^2_x &=   \eta k^2\chi_{\perp} \left[ 1- A \right]  , \label{eqA3} \\
\omega^2_y &= \omega^2_z =   \frac{2\eta\chi_{\perp}}{w^2_0}\left[ 1- A \right] , \label{eqA4}  \\
\omega^2_{\theta_y} &= \frac{4\eta}{a^2} \left[  \Delta\chi \left( 1-2A \right)  -  \frac{k^2a^2\chi_{\perp}}{8}  \right] , \label{eqA6} \\
\omega^2_{\theta_z} &= \frac{\eta k^2\chi_{\perp}}{2}  , \label{eqA7} \\
\omega^2_1 &= \eta  \left[ 2k^2A\chi_{\perp} - \frac{\Delta\chi_{\perp}}{w^2_0} \right], \label{eqA8} \\
\omega^2_2 &= 2\eta k^2 A\chi_{\perp}, \label{eqA9}
\end{align}
where the common factor $\eta = \frac{2\epsilon_0E^2_0}{\rho}$ with $\rho$ the mass density, the moment of inertia for negligible thickness $I_y=I_x=ma^2/4$ was used, and $A= a^2/4w^2_0$ is one quarter the square ratio of the radius to the waist. For calculations of the frequencies in the main text the relation $E^2_0 = 4P/(\epsilon_0c\pi w^2_0)$ is used where $P$ is the total laser power.

In experiments the beam waists in the $z$ and $y$ directions are often not symmetric. The asymmetric Gaussian leads the frequencies above to be altered slightly and can be accounted for by using different beam waists in Eq. \eqref{eqA2}.

 One noteworthy feature not mentioned in the main text is that the $\omega_{\theta_z}$ rotational frequency depends on $\chi_{\perp}$ rather than $\Delta\chi$. Rotational frequencies in the Rayleigh approximation depend on $\Delta\chi$ solely \cite{Seberson2019} as the particle's long axis tries to align with the electric field. In the Rayleigh approximation $\omega_{\theta_z}=0$ and is only non-zero here due to the electric field gradient across the finite extension of the disk.

\bibliographystyle{apsrev4-1}
\bibliography{levitated_disks}

\end{document}